\documentclass[12pt]{iopart}

\usepackage{iopams}
\usepackage{setstack}

\usepackage{epsfig}
\usepackage{graphics}

\eqnobysec

\begin{document}

\title{Expressing Decoherence with Spectral and Stochastic Methods}

\author{D Salgado and JL S\'{a}nchez-G\'{o}mez}
\address{Dpto. F\'{\i}sica Te\'{o}rica, Universidad Aut\'{o}noma de Madrid\\
28049 Cantoblanco, Madrid (Spain)}

\eads{\mailto{david.salgado@uam.es},\mailto{jl.sanchezgomez@uam.es}}

\begin{abstract}
We suggest a novel proposal to express decoherence in open quantum systems by jointly employing spectral and stochastic methods. This proposal, which basically perturbs the unitary evolution operator in a random fashion, allows us to embrace both markovian and nonmarkovian situations with little extra effort. We argue that it can be very suitable to deal with models where an approximation neglecting some degrees of freedom is undertaken. Mathematical simplicity is also obtained both to solve some master equations and to arrive at experimentally measured decoherence functions.  
\end{abstract}

\pacs{03.65.Yz,02.50.-r}

\maketitle

\section{Introduction}
Decoherence in an open quantum system is the loss of quantum coherence due to the entanglement with its environment as a consequence of their mutual interaction \cite{GiuJooKieKupStaZeh96a}. It precludes thus the formation of some superpositions of states, creating in the system environment-induced superselection rules \cite{Zur82a,Zur91a}. The rigorous mathematical expression of this phenomenon runs parallel to this scheme, namely, one should first consider the joint evolution of both the system and the environment and then take the partial trace over the environmental degrees of freedom \cite{Dav76a,BrePet02a}. In general, the resulting evolution equation for the density operator of the system $\rho_{\rm S}$ is non-markovian and extremely difficult to handle. This is the reason why suitable approximations have been traditionally pursued. In particular, adequate methods to derive an approximate \emph{markovian} equation have been found \cite{BrePet02a,GorFriVerKosSud78a,AliLen87a}. On the other hand, more axiomatic results have also been developed which pose some general physical conditions for the evolution of the system density operator \cite{Lin76a,GorKosSud76a} and arrive at the general form for the evolution equation of $\rho_{\rm S}$. Both approaches show pros and cons: the axiomatic program starts by assuming the markovianity of the evolution as one of the physical hypotheses, thus making rather difficult the generalization to non-markovian situations, which, due to experimental progress, are becoming interesting enough to develop theoretical tools. On the other hand, the constructive approach depends very sensitively on the particular model of interaction between the system and its environment, thus rendering it rather inappropiate for generic discussion beyond precise details.\\
Here we suggest another proposal which embraces both markovian and non-markovian situations, like the constructive approach, but it does not depend, in general terms, on the specific features of the system-environment interaction. The physical picture which motivates this scheme models the effect of the environment upon the system by a random perturbation of the evolution operator. Since the origin of the stochasticity is rooted upon the environment and since we do not have control over its degrees of freedom, we then take the stochastic expectation value. This is of course not new, and indeed it has been used in classical physics too \cite{Gar85a,Kam81a}. The novel proposal stems from the combination of this old idea with the spectral decomposition theorem for the evolution operator \cite{Kre78a,DunSch63b}. This combination will turn out to be quite useful to express decoherence in certain models.\\
The paper is organized as follows. In section \ref{GenSch} the general scheme of this proposal is presented, then in section \ref{GloPer} the case of global random perturbations of the Hamiltonian spectrum is studied to arrive at a Lindblad-type evolution equation both when the Lindblad operators are selfadjoint (subsection \ref{SelLin}) and nonselfadjoint (subsection \ref{NonLin}). In the following section (\ref{LocPer}) a detailed example of the conjuction of the spectral decomposition and stochastic methods is worked out. A brief discussion and some conclusions are then presented in section \ref{DisCon}. Finally we close the paper with a computational appendix (\ref{Com}) and a mathematical proof (\ref{InitCorr}).

\section{General scheme\label{GenSch}}
To show how stochastic methods and the spectral decomposition theorem are combined, consider a quantum system with hamiltonian $H$, which for simplicity's sake will be discrete and nondegenerate. The evolution equation for this system will then be given by the evolution operator $U(t)=\exp(-\rmi tH)$, which by means of the spectral decomposition of $H=\sum_{k}\epsilon_{k}P_{k}$ can be expressed as ($\hbar=1$ as usual) 

\begin{equation}
U(t)=\sum_{k}e^{-\rmi \epsilon_{k}t}P_{k}
\end{equation}

The operator $U(t)$ will then be randomly perturbed as the following relation shows

\begin{equation}\label{StocProm}
U(t)=\sum_{k}e^{-\rmi \epsilon_{k}t}P_{k}\to U_{st}(t)=\sum_{k}e^{-\rmi \chi_{k}(t)}P_{k}
\end{equation}

\noindent where $\chi_{k}(t)$ denotes a real-valued stochastic process for each $k$. The system density operator $\rho_{\rm S}(t)$ at time $t$ will then be given by the expectation value $\rho_{\rm S}(t)=\mathbb{E}[U_{st}(t)\rho(0)U_{st}^{\dagger}(t)]$. The density operator, in the Hamiltonian eigenvector basis, is then given by

\begin{equation}
\rho_{n,m}(t)=\mathbb{E}[e^{-\rmi (\chi_{n}(t)-\chi_{m}(t))}]\rho_{n,m}(0)
\end{equation}

\noindent where it has been supposed that the system and the environment are not initially correlated, thus $\rho(0)$ is not random. Both the original unitary evolution and the decoherence function are contained in the expectation value $\mathbb{E}[e^{-\rmi(\chi_{n}(t)-\chi_{m}(t))}]$ and should then be properly disentangled. To do this note that the ``stochastic promotion'' expressed by \eref{StocProm} can be very intuitively understood if we express the argument of the exponential in the original evolution operator as an integral and the stochastic process $\chi_{k}(t)$ is written as an Ito process \cite{Oks98a}

\begin{equation}\label{StocPromIto}
e_{k}t=\int_{0}^{t}e_{k}ds\to\chi_{k}(t)=\int_{0}^{t}a_{k}(s)ds+\int_{0}^{t}b_{k}(s)d\mathcal{B}_{k}(s)
\end{equation}

\noindent where $a_{k}(t)$ and $b_{k}(t)$ are real-valued stochastic processes for each $k$ and $\mathcal{B}_{k}(t)$ are standard real Brownian motions for each $k$.\\
Under full generality the previous expectation value cannot be computed, but with some general physical assumptions some useful results may rapidly arise. As a very elementary example consider for instance the white-noise perturbation, i.e.

\begin{equation}\label{WhiNoi}
e_{k}t\to e_{k}t+\int_{0}^{t}e_{k}\xi(s)ds=e_{k}t+e_{k}\mathcal{B}_{k}(t)
\end{equation}

\noindent where $\xi(t)$ denotes white noise with correlation properties $\mathbb{E}[\xi(t)]=0$, $\mathbb{E}[\xi(t)\xi(s)]=\gamma\delta(t-s)$. Note that all eigenvalues are perturbed by the same stochastic noise $\xi(t)$. Under this assumption and after computing the expectation value $$\mathbb{E}[e^{-\rmi (e_{k}-e_{k'})\int_{0}^{t}\xi(s)ds}]=e^{-\frac{\gamma t}{2}(e_{k}-e_{k'})^{2}}$$ the elements of $\rho_{\rm S}(t)$ in the Hamiltonian eigenvector basis are given by

\begin{equation}\label{CompDensNorm}
\rho_{n,m}(t)=\rho_{n,m}^{(0)}(t)e^{-\frac{\gamma t}{2}(e_{n}-e_{m})^{2}}
\end{equation}

\noindent where $\rho^{(0)}(t)$ denotes the unitarily evolved density operator. The corresponding evolution equation for $\rho_{\rm S}$ is straightforwardly shown to be the markovian phase-destroying master equation (see \cite{SalSan02d} for an alternative deduction)

\begin{equation}\label{PDME}
\frac{d\rho_{\rm S}(t)}{dt}=-\rmi[H,\rho_{\rm S}(t)]-\frac{\gamma}{2}[H,[H,\rho_{\rm S}(t)]]
\end{equation}

Thus the effect of the environment can be understood as a \emph{global} random energy ``kick'' upon the system. This method also provides a very fast way to solve equation \eref{PDME} provided one knows the solution to the unitary evolution (cf. \cite{SalSan02b} for an explicit example).


\section{Global perturbations\label{GloPer}}

\subsection{Selfadjoint Lindblads\label{SelLin}}
The \emph{global} random perturbations like in the example in the preceding section can also be expressed without resorting to the spectral decomposition theorem. A partial treatment was provided in \cite{SalSan02b}, where the most general Lindbladian evolution with selfadjoint Lindblad operators was obtained. Here we offer a complementary proof of this result which uses different techniques and thus paves the way for later generalizations (nonselfadjoint Lindblad operators). Let $H$ be the Hamiltonian of the system. Introduce now a global random perturbation by adding the operator $\eta(t)V$ to $H$, where $\eta(t)$ denotes an arbitrary real-valued stochastic process and $V$ an arbitrary selfadjoint operator. The evolution operator will then read

\begin{equation}\label{RanEvolOp}
U_{st}(t)=\exp[-\rmi tH-\rmi\eta(t)V]
\end{equation}

The main trouble to compute $\rho(t)\equiv\mathbb{E}[U_{st}(t)\rho(0)U_{st}^{\dagger}(t)]$ stems from the commutativity properties of $H$ and $V$. If $[H,V]=0$ it can be readily shown \cite{SalSan02b} that the evolution equation for $\rho_{\rm S}$ is

\begin{equation}
\frac{d\rho_{\rm S}(t)}{dt}=-\rmi[H,\rho_{\rm S}(t)]-\frac{1}{2}\frac{d}{dt}\langle\eta\rangle_{t}[V,[V,\rho_{\rm S}(t)]]
\end{equation}

\noindent where $\langle\eta\rangle_{t}$ denotes the quadratic variation of $\eta(t)$ \cite{KarShr91a}. When $[H,V]\neq 0$, the computation cannot be performed in the same way. An option to overcome this difficulty was presented in \cite{SalSan02b}, where the stochastic perturbation was performed in the Heisenberg picture; here we resort to Feynman's operational calculus \cite{Fey51a,dFacJohLap97a,JohLap00a}. In particular, the following three Feynman's heuristic rules to deal with functions of noncommuting operators will be thoroughly used:

\begin{enumerate}
\item\label{Rule1} Attach time indices to the operators to specify the order of operators in products.
\item\label{Rule2} With time indices attached, form functions of these operators by treating them as though they were commuting.
\item\label{Rule3} Finally, ``disentangle'' the resulting expressions; that is, restore the conventional ordering of the operators.
\end{enumerate}

A mathematical formalization of each of these rules can be found in \cite{dFacJohLap97a,JohLap00a}. Here we will only show how they can be used to prove our result. Consider then the evolution operator \eref{RanEvolOp}; attach time indices both to $H$ and $V$ so that one can write

\numparts
\begin{equation}
U_{st}(t)=\exp\left(-\rmi\int_{0}^{t}H(s)ds-\rmi\int_{0}^{t}V(s)d\mu(s)\right)
\end{equation}

\noindent where $d\mu(t)$ denotes the Lebesgue-Stieltjes measure $\eta(t)dt$. Now treat both factors as though they were commuting:

\begin{equation}
  U_{st}(t)=\exp\left(-\rmi\int_{0}^{t}H(s)ds\right)\exp\left(-\rmi\int_{0}^{t}V(s)d\mu(s)\right)
\end{equation}

The density operator then reads

\begin{equation}
\fl\rho(t)=e^{-\rmi\int_{0}^{t}H(s)ds}\mathbb{E}\left[\exp\left(-\rmi\int_{0}^{t}V(s)d\mu(s)\right)\rho(0)\exp\left(\rmi\int_{0}^{t}V(s)d\mu(s)\right)\right]e^{\rmi\int_{0}^{t}H(s)ds}
\end{equation}

Now the expectation value can be computed using (see also appendix \ref{Com}) $\rme^{-\rmi A}B\rme^{\rmi A}=\rme^{-\rmi [A,\cdot]}B$ with $A=A^{\dagger}$, where $\rme^{-\rmi [A,\cdot]}B\equiv\sum_{n=0}^{\infty}\frac{(-\rmi)^{n}}{n!}[A,[A,[A,\stackrel{n}{\dots},[A,B]]]]$:

\begin{equation}
\fl\mathbb{E}\left[\exp\left(-\rmi\int_{0}^{t}V(s)d\mu(s)\right)\rho(0)\exp\left(\rmi\int_{0}^{t}V(s)d\mu(s)\right)\right]=\exp\left(-\frac{1}{2}\left[\int_{0}^{t}V^{2}(s)d\langle\eta\rangle_{s},\cdot\right]\right)\rho(0)
\end{equation}

And finally, restore the conventional ordering of the operators:

\begin{equation}
\rho_{\rm S}(t)=\exp\left(-\rmi[H,\cdot]-\frac{1}{2}\langle\eta\rangle_{t}[V^{2},\cdot]\right)\rho(0)
\end{equation}

\noindent which drives us to the master equation

\begin{equation}
\frac{d\rho_{\rm S}(t)}{dt}=-\rmi[H,\rho_{\rm S}(t)]-\frac{1}{2}\frac{d}{dt}\langle\eta\rangle_{t}[V,[V,\rho_{\rm S}(t)]]
\end{equation}
\endnumparts

We have proved the main result found in \cite{SalSan02b} with other means: \emph{Any Lindbladian master equation, \textbf{either markovian or non-markovian}, with selfadjoint Lindblad operators, can be understood as a random unitary evolution}. The generalization to many Lindblad operators can be readily accomplished by using several uncorrelated stochastic processes $\eta_{i}(t)$ $i=1,\dots,n$. Note that the markovian case is obtained only when $\langle\eta\rangle_{t}$ is linear in time, i.e. when $\eta(t)=\rm{const.}+\gamma^{1/2}\mathcal{B}_{t}$ which amounts to perturbing the original Hamiltonian by a white noise like in \eref{WhiNoi}.

\bigskip

Complementarily one can also use these techniques to study the evolution of a system given by a Hamiltonian with one or several stochastic parameters in it. Take for instance the system with Hamiltonian $H=H_{1}+\lambda(t)H_{2}$ and let the parameter $\lambda(t)$ be uncontrollably random. Following the previous prescriptions the corresponding evolution operator will read

\begin{equation}
U_{st}(t)=\mathcal{T}\exp\left[-\rmi tH_1-\rmi H_{2}\left(\mu(t)+\int_{0}^{t}\tilde{\lambda}(s)d\mathcal{B}(s)\right)\right]
\end{equation}

\noindent where $\mathcal{T}$ denotes the time-ordering operator and by definition $\int_{0}^{t}\lambda(s)ds\equiv\mu(t)+\int_{0}^{t}\tilde{\lambda}(s)d\mathcal{B}(s)$. The master equation for $\rho_{\rm S}$ will then be

\begin{equation}
\fl\frac{d}{dt}\rho_{\rm S}(t)=-\rmi[H_{1},\rho_{\rm S}(t)]-\rmi\dot{\mu}(t)[H_{2},\rho_{\rm S}(t)]-\frac{1}{2}\frac{d}{dt}\langle\mu(t)+\int_{0}^{t}\tilde{\lambda}(s)d\mathcal{B}(s)\rangle_{t}[H_{2},[H_{2},\rho_{\rm S}(t)]]
\end{equation}

This scheme has already been used \cite{SchneiMil97a,SalSan02b} to study the effects of laser intensity and phase fluctuations in ion traps. In more general terms, these techniques appear very suitable to deal with approximate models in which the approximation renders the physical realistic situation mathematically manageable to perform computations, but on the other hand it ineludibly carries some decohering effect in it. An explicit example was worked out in \cite{SalSan02b}, where the Jaynes-Cummings model (JCM hereafter) was revisited within this approach. The JCM is a crude, though very appropiate, approximation to deal with the interaction between an atom and the electromagnetic field in certain circumstances (cf. \cite{JayCum63a,ShoKni93a}) which reduces both systems to a two-level system and a harmonic oscillator, respectively, interacting via a Hamiltonian $H_{I}=\lambda (\sigma_{-}a^{\dagger}+\sigma_{+}a)$, where $\lambda$ is the coupling constant, $\sigma_{\pm}$ are the raising/lowering atomic operators and $a(a^{\dagger})$ are the annihilation (creation) field mode operators. It is then clear that the rest of field modes as well as other atomic degrees of freedom are completely neglected, thus providing an ideal situation to quantitatively describe some physical phenomena like Rabi oscillations, periodic collapses and revivals or squeezing \cite{PenLi98a}. Decoherence is absent in this model, however experimental results \cite{MeeMonKinItaWin96a,BruSchmMaaDreHagRaiHar96a} show the opposite, though extremely difficult to be attributed to the environment. This tiny decoherence effect may be incorporated into the model from the beginning by letting the coupling constant $\lambda$ have a random character, this randomness coming from the effect of the neglected degrees of freedom. The theoretical predictions are in good agreement with experimental results \cite{SalSan02b}. Note that within this approach decoherence should be understood as an \emph{intrinsic} phenomenon stemming out from the adopted approximation and thus being an inherent property in those models which neglects degrees of freedom of the system. Notice how despite providing the same analytical behaviour, this approach conceptually differs from others in which a quantum bosonic reservoir is explicitly taken into account \cite{MurKni98a,BosKniMurPleVed98a}.

\subsection{Nonselfadjoint Lindblads\label{NonLin}}
A natural question arises concerning whether it is possible or not ot arrive at a Lindbladian master equation with Lindblad operators not necessarily selfadjoint like in the previous examples. And the answer is positive but with important remarks. To be concrete let us consider a spin-$1/2$ interacting with a random magnetic field $\mathbf{B}(t)=B_{\rm x}(t)\mathbf{\hat{x}}+B_{\rm y}(t)\mathbf{\hat{y}}+B_{\rm z}(t)\mathbf{\hat{x}}$. The Hamiltonian describing the evolution of the spin will then be 

\begin{equation}\label{QBitHamilt}
H(t)=\omega_{0}\sigma_{\rm z}+\mathbf{B}(t)\cdot\bsigma
\end{equation}

Complementarily this can be viewed as a quantum bit (left term of the rhs of \eref{QBitHamilt}) undertaking a logical operation in a noisy environment (right term of the rhs). Without loss of generality we can focus on those cases in which $\mathbf{B}(t)\cdot\mathbf{z}=0$, since otherwise we can use the tools developed in the preceding section. The evolution operator will then be

\numparts
\begin{equation}
U_{st}(t)=\mathcal{T}\exp\left[-\rmi\omega_{0}t-\rmi\int_{0}^{t}\left(B_{\rm x}(s)\sigma_{\rm x}+B_{\rm y}(s)\sigma_{\rm y}\right)ds\right]
\end{equation}

Since we will make use of Feynman's operational calculus as in previous paragraphs, we can concentrate on the operator

\begin{equation}\label{NewQOp}
\tilde{U}_{st}(t)=\mathcal{T}\exp\left[-\rmi\int_{0}^{t}\left(B_{\rm x}(s)\sigma_{\rm x}(s)+B_{\rm y}(s)\sigma_{\rm y}(s)\right)ds\right]
\end{equation}
\endnumparts

The magnetic field components will be expressed through Ito integrals as 

\numparts
\begin{eqnarray}
\int_{0}^{t}B_{\rm x}(s)ds&\equiv&\int_{0}^{t}b_{\rm x}(s)d\mathcal{B}_{\rm x}(s)\\
\int_{0}^{t}B_{\rm y}(s)ds&\equiv&\int_{0}^{t}b_{\rm y}(s)d\mathcal{B}_{\rm y}(s)
\end{eqnarray}
\endnumparts

\noindent where the $b_{k}(t)$'s are arbitrary deterministic real-valued functions and $\mathcal{B}_{k}(t)$ denotes nonstandard real Brownian motions (cf. below). The careful reader may argue at first sight that this is not the more general form for a real stochastic process, and it is not, though it contains all physically relevant cases which interest us in this analysis. To prove that, we begin by demanding that the magnetic field components be martingales \cite{Oks98a,KarShr91a}, which is general enough from a physical standpoint. Then making use of the martingale representation theorem, they can be uniquely written as

\begin{equation}
\int_{0}^{t}B_{k}(s)ds=\mathbb{E}\left[\int_{0}^{t}B_{k}(s)ds\right]+\int_{0}^{t}g_{k}(s)d\mathcal{B}_{k}(s)\quad k=\rm x,\rm y
\end{equation} 

\noindent where $g_{k}(t)$ is a real stochastic process depending on the particular choice of $B_{k}(t)$. Without loss of generality we take $\mathbb{E}[B_{k}]=0$ since otherwise this factor can be included in the dropped-out exponential after application of Feynman's rules and finally recovered back in the final step. We are interested in those effects coming out from the stochasticity of the Hamiltonian and which are at the end detected after calculating the stochastic average. This is the reason why we can also consider the processes $g_{k}(t)$'s as being deterministic.\\
The correlations properties of the magnetic field components will be generically expressed through the correlation properties of the Brownian motions

\numparts
\begin{eqnarray}
\mathbb{E}[\mathcal{B}_{\rm x}^{2}(t)]&=&\gamma_{\rm x}t\\
\mathbb{E}[\mathcal{B}_{\rm y}^{2}(t)]&=&\gamma_{\rm y}t\\
\mathbb{E}[\mathcal{B}_{\rm x}(t)\mathcal{B}_{\rm y}(t)]&=&\gamma_{\rm xy}t
\end{eqnarray}
\endnumparts

In order to expouse clearly the role played by the correlation properties between both componentes, let us concentrate upon white-noise process, i.e. $b_{\rm x}(t)=b_{\rm y}(t)=1$ (equivalently $B_{\rm x}(t)=B_{\rm y}(t)=\xi(t)$). Now we decompose $\sigma_{\rm x}$ and $\sigma_{\rm y}$ in terms of the raising and lowering operators to rewrite \eref{NewQOp} as 

\numparts
\begin{equation}\label{URaisLow}
\tilde{U}_{st}(t)=\mathcal{T}\exp\left[-\rmi\int_{0}^{t}\left(B(s)\sigma_{\rm +}(s)+B^{*}(s)\sigma_{\rm -}(s)\right)ds\right]
\end{equation}

\noindent where $B(t)$ is a \emph{complex} stochastic process with $\mathfrak{R}B(t)=B_{\rm x}(t)$ and $\mathfrak{I}B(t)=-B_{\rm y}(t)$. Alternatively equation \eref{URaisLow} can be rewritten as

\begin{equation}
\tilde{U}_{st}(t)=\mathcal{T}\exp\left[-\rmi\int_{0}^{t}\sigma_{+}(s)d\mathcal{B}(s)-\rmi\int_{0}^{t}\sigma_{-}(s)d\mathcal{B}^{*}(s)\right]
\end{equation}
\endnumparts

\noindent where $\mathcal{B}(t)$ is a complex stochastic process with correlation properties given by

\numparts
\begin{eqnarray}
\mathbb{E}[B^{2}(t)]&=&\left(\gamma_{\rm x}-\gamma_{\rm y}-\rmi2\gamma_{\rm xy}\right)t\\
\mathbb{E}[B(t)B^{*}(t)]&=&(\gamma_{\rm x}+\gamma_{\rm y})t
\end{eqnarray}
\endnumparts

Now using again $\rme^{-\rmi A}B\rme^{\rmi A}=\rme^{-\rmi [A,\cdot]}B$ with $A=A^{\dagger}$ we compute the stochastic expectation value $\tilde{\rho}(t)\equiv\mathbb{E}\left[\tilde{U}_{st}(t)\rho(0)\tilde{U}_{st}^{\dagger}\right]$:

\begin{equation}
\tilde{\rho}(t)=\mathcal{T}\mathbb{E}\left[\exp\left(-\rmi\int_{0}^{t}[\sigma_{+}(s),\cdot]d\mathcal{B}(s)-\rmi\int_{0}^{t}[\sigma_{-}(s),\cdot]d\mathcal{B}^{*}(s)\right)\right][\rho(0)]
\end{equation} 

After computing this expectation value (see appendix \ref{Com}) and restoring the conventional time ordering of operators, the final result is (up to Hamiltonian factors)

\begin{equation}\label{Result}
\tilde{\rho}(t)=\exp\left[t\left(\mathcal{L}_{\rm{xy}}+\mathcal{L}_{\rm +}+\mathcal{L_{\rm -}}+\mathcal{D}_{\rm{xy}}\right)\right][\rho(0)]
\end{equation}

\noindent where 

\numparts
\begin{eqnarray}
\mathcal{L}_{\rm{xy}}&\equiv&-\frac{\gamma_{\rm x}-\gamma_{\rm y}}{2}\left([\sigma_{\rm x},[\sigma_{\rm x},\cdot]]-[\sigma_{\rm y},[\sigma_{\rm y},\cdot]]\right)\\
\label{DissLind}\mathcal{L}_{\rm j}&\equiv&(\gamma_{\rm x}+\gamma_{\rm y})\left([\sigma_{\rm j}\cdot,\sigma_{\rm j}^{\dagger}]+[\sigma_{\rm j},\cdot\sigma_{\rm j}^{\dagger}]\right)\quad j=+,-\\
\mathcal{D}_{\rm{xy}}&\equiv&2\gamma_{\rm{xy}}\left(\sigma_{\rm x}\cdot\sigma_{\rm y}+\sigma_{\rm y}\cdot\sigma_{\rm x}\right)
\end{eqnarray}
\endnumparts

The first two generators are Lindblad generators, thus denoting a completely positive markovian evolution \cite{Lin76a,GorKosSud76a} whereas the last one is not. Since the evolution is clearly markovian, this means that complete positivity is not ensured during the evolution. It is remarkable how once more the presence of correlations spoils the complete positivity \cite{SteBuz01a,Pec94a}. In fact this is a clear example of how correlations \emph{within the environment}, even in the case of no correlations between the system and this environment, prevents the evolution from being completely positive (see \ref{InitCorr} for a mathematical proof of this fact). Notice however that when the random magnetic field is isotropic ($\gamma_{\rm x}=\gamma_{\rm y}$) and the components are uncorrelated ($\gamma_{\rm{xy}}=0$), a fully Lindblad evolution is recovered with dissipative generator given by \eref{DissLind}, i.e. a Lindblad evolution with nonselfadjoint Lindblad operators. Nevertheless this generator shows two opposite parts which within this approach cannot be separated. Physically this is rooted upon the isotropy of the magnetic field which, as expected, drives the system with equal probability both to the ground and to the excited state.\\
The generalization to nonmarkovian evolution can be performed with little extra effort. For simplicity we keep the isotropy condition, thus $b_{\rm x}(t)=b_{\rm y}(t)\equiv b(t)$. The random evolution operator will then be

\begin{equation}
\tilde{U}_{st}(t)=\mathcal{T}\exp\left(-\rmi\int_{0}^{t}b(s)\sigma_{\rm +}(s)d\mathcal{B}(s)-\rmi\int_{0}^{t}b(s)\sigma_{\rm -}(s)d\mathcal{B}^{*}(s)\right)
\end{equation}

\noindent where the complex stochastic process $\mathcal{B}(t)$ have the same correlation properties as above. The calculation proceeds along similar lines with result

\begin{equation}\label{ResultNonMark}
\tilde{\rho}(t)=\exp\left[\Lambda(t)\left(\mathcal{L}_{\rm{xy}}+\mathcal{L}_{\rm +}+\mathcal{L_{\rm -}}+\mathcal{D}_{\rm{xy}}\right)\right][\rho(0)]
\end{equation}

\noindent where the generators are the same as before and $\Lambda(t)\equiv\int_{0}^{t}b^{2}(s)ds$. This is clearly nonmarkovian and reduces to the markovian case only when $b(t)=\rm{const}$. Regretfully the previous calculations show that the dissipative generator appears always in opposite pairs. The conditions under which a single dissipative Lindblad generator is obtained are being currently under study (see also \cite{SalSan02c} for an alternative approach).

\section{Local perturbations\label{LocPer}}

But the versatility of the spectral decomposition theorem appears when the perturbation does not affect the whole system globally, as in the previous examples. By this we mean that in the random perturbation \eref{StocPromIto} the Brownian motions $\mathcal{B}_{k}(t)$ do not show singular correlation properties\footnote{Note that in the previous cases --global perturbations-- the processes $\mathcal{B}_{k}(t)$'s satisfied $\mathbb{E}[\mathcal{B}_{k}(t)\mathcal{B}_{k'}(t)]=t$ for all pairs $k,k'$, i.e. all $\mathcal{B}_{k}(t)$'s were esentially copies of the same standard one $\mathcal{B}(t)$ in \eref{WhiNoi}.}, but rather general ones. Take for instance the case in which 

\begin{equation}
e_{k}t\to e_{k}t+\gamma e_{k}\mathcal{B}_{k}(t)
\end{equation}

\noindent with correlation properties $\mathbb{E}[\mathcal{B}_{k}(t)\mathcal{B}_{k'}(t)]=\delta_{kk'}t$. Under this assumption (which physically can be understood as the independent or incoherent effect of the environment upon each system energy level), the density operator components are given by

\numparts
\begin{eqnarray}
\label{CompDensnnUNCORR}\rho_{nm}(t)&=&\rho_{nn}^{(0)}(t)\quad n=m\\
\label{CompDensnmUNCORR}&=&\rho_{nm}^{(0)}(t)e^{-\frac{\gamma t}{2}(e^{2}_{n}+e^{2}_{m})}\quad n\neq m
\end{eqnarray}
\endnumparts

\noindent which show a faster decoherence than \eref{CompDensNorm}. By choosing another correlation properties among the $\mathcal{B}_{k}(t)$'s one can arrive at more general decoherence functions for the decay of off-diagonal terms of the density operator. As a prominent example let us consider two quantum particles interacting through a central potential $V(r_{12})$. The Hamiltonian will then be given by 

\begin{equation}
H=H_{1}^{0}+H_{2}^{0}+H_{12}=-\frac{1}{2m_{1}}\Delta_{1}-\frac{1}{2m_{2}}\Delta_{2}+\lambda V(r_{12})
\end{equation}

\noindent where $\lambda$ is a coupling constant and $r_{12}\equiv|\mathbf{r}_{1}-\mathbf{r}_{2}|$. In terms of the center-of-mass $\mathbf{R}_{CM}\equiv\mathbf{R}$ and the relative position $\mathbf{r}_{12}\equiv\mathbf{r}$ coordinates, the Hamiltonian is given by

\begin{equation}
H=-\frac{1}{2M}\Delta_{\rm{CM}}-\frac{1}{2\mu}\Delta_{\rm{rel}}+\lambda V(r)
\end{equation}

\noindent with $M$ and $\mu$ the total and  reduced masses respectively. The total energy is written as 

\begin{equation}
E_{T}=E_{CM}+E_{rel}
\end{equation}

\noindent so that the corresponding evolution operator will then be given by

\begin{equation}\label{EvolOpGen}
U(t)=\int_{\sigma(H_{\rm{CM}})\times\sigma(H_{\rm{rel}})}e^{-\rmi tE_{T}}dP(E_{T})
\end{equation}

Now a weak environment-system coupling can be expressed through a random perturbation of the total energy $E_{T}$ by a stochastic amount generically expressed like $\int e(s;E_{T})d\mathcal{B}(s;E_{T})$ where $e(s;E_{T})\equiv\bar{e}(s;E_{\rm{CM}})$ only depends on the center of mass and where the standard real Brownian motions show correlation properties only dependent on the center of mass variables as well. This is the main physical hypothesis and it is expressed as follows

\numparts
\begin{eqnarray}
U_{st}(t)&=&\int_{\sigma(H_{\rm{CM}})\times\sigma(H_{\rm{rel}})}e^{-\rmi tE_{T}-\rmi\int_{0}^{t}e(s;E_{T})d\mathcal{B}(s;E_{T})}dP(E_{T})\\
&=&\int_{\sigma(H_{\rm{CM}})\times\sigma(H_{\rm{rel}})}e^{-\rmi tE_{T}-\rmi\int_{0}^{t}\bar{e}(s;E_{\rm{CM}})d\mathcal{B}(s;E_{T})}dP(E_{T})
\end{eqnarray}
\endnumparts

\noindent with $\mathbb{E}[\mathcal{B}(t;E_{T})\mathcal{B}(t;E'_{T})]=g(t;E_{T},E'_{T})$. Note that the correlation function satisfies

\numparts
\begin{eqnarray}
g(t;E_{T},E_{T})&=&1\\
g(t;E_{T},E'_{T})&\leq&1\\
\label{CorrelPropCMREL}g(t,E_{T},E'_{T})&=&\bar{g}(t;E_{\rm{CM}},E'_{\rm{CM}})
\end{eqnarray}
\endnumparts

\noindent where the last equation reflects the assumed physical hypothesis. Then if we denote the stochastic process $$\eta(t;E_{T},E'_{T})\equiv\int_{0}^{t}e(s;E_{T})d\mathcal{B}(s;E_{T})-\int_{0}^{t}e(s;E'_{T})d\mathcal{B}(s;E'_{T})$$the components of the density operator in the basis $|E_{\rm{CM}};E_{\rm{rel}}\rangle$ will then be given by

\begin{equation}
\langle E_{\rm{CM}};E_{\rm{rel}}|\rho(t)|E'_{\rm{CM}};E'_{\rm{rel}}\rangle=\rho^{(0)}(E_{\rm{CM}},E_{\rm{rel}};E'_{\rm{CM}},E'_{\rm{rel}})e^{-\frac{1}{2}\langle\eta(\cdot;E_{T},E'_{T})\rangle_{t}}
\end{equation}

Now the quadratic variation of $\eta$ can be calculated in terms of the correlation function $g$ above:

\begin{eqnarray}
\fl\langle\eta(\cdot;E_{T},E'_{T})\rangle_{t}&=&\int_{0}^{t}\left[e^{2}(s;E_{T})+e^{2}(s;E'_{T})-2g(s;E_{T},E'_{T})e(s;E_{T})e(s;E'_{T})\right]ds=\nonumber\\
&\lo=&\int_{0}^{t}\left[\bar{e}^{2}(s;E_{\rm{CM}})+\bar{e}^{2}(s;E'_{\rm{CM}})-2\bar{g}(s;E_{\rm{CM}},E'_{\rm{CM}})\bar{e}(s;E_{\rm{CM}})\bar{e}(s;E'_{\rm{CM}})\right]ds\nonumber\\
&\lo\equiv&\langle\bar{\eta}(\cdot;E_{\rm CM},E'_{\rm CM})\rangle_{t}
\end{eqnarray}

Note that $\langle\eta(\cdot;E_{T},E_{T})\rangle_{t}=0$, thus the trace is conserved.  Also notice that components showing quantum coherence for different center-of-mass modes asymptotically vanish, since 

\begin{equation}
\frac{d}{dt}\langle\eta(\cdot;E_{T},E'_{T})\rangle_{t}\geq(\bar{e}(s;E_{\rm{CM}})-\bar{e}(s;E'_{\rm{CM}}))^{2}\geq 0
\end{equation}

On the other hand, full quantum coherence is kept for the relative coordinate, thus internal dynamics is not affected. A particular example of this scheme has been applied to arrive at the experimental expressions obtained in the study of Rabi oscillations in an ion trap \cite{SalSan02b}.\\
This same technique can also be applied in those cases in which the randomness appears only in one part of the total Hamiltonian. Let for instance $H=H_{0}+H_{\rm{I}}$ be the total Hamiltonian of a quantum system and suppose that the randomness affects only the interaction part of $H$, i.e. $H_{\rm{I}}$. One has two complementary alternatives. On one hand, one may change to the $H_{0}$-picture and then apply the preceding techniques to the evolution operator $U_{\rm{int}}(t)=\mathcal{T}\exp\left(-\rmi\int_{0}^{t}H_{\rm{I}}(s)ds\right)$ where $H_{\rm{I}}(t)$ denotes the $H_{\rm{I}}$ Hamiltonian in $H_{0}$ representation, i.e. $H_{\rm{I}}(t)=\rme^{\rmi tH_{0}}H_{\rm{I}}\rme^{-\rmi tH_{0}}$. On the other hand, one may also resort to Feynman's operational calculus and ``break'' the exponential into two parts: a first deterministic exponential involving only $H_{0}$ and a second one involving the random interaction Hamiltonian $H_{I}$, which can now be expressed through the spectral decompostion theorem and then stochasticly promoted like above. Both approaches are equivalent. 

\section{Discussion and conclusions\label{DisCon}}

The use of stochastic methods in Hilbert space is of course not new (cf. \cite{BrePet02a,DioLuc94a} and references therein). Here they have been used in conjuction with the spectral decomposition to accomodate a great variety of decoherence functions which may appear in realistic situations and which otherwise require more or less complex calculations. It is important to remark that the whole scheme presented above preserves the unitarity of the evolution operator, i.e. $U_{st}(t)$ is unitary with probability $1$. This is in contrast to the radically different usage of these methods in stochastic dynamical reduction theories (cf. \cite{AdlBroBruHug01a} and references therein for diverse examples within this philosophy), where the norm of the initial state vector is not preserved (reduction) and thus one is obliged to introduce non-linear terms to keep the probability interpretation. A clear-cut example of the difference between the decoherence and the reduction behaviours appears in \cite{Spi94a}. Possible connections between these two distinct pictures (decoherence and reduction) are currently under study. We argue that these possible connections, if any, appear in those cases where an analytical continuation in the complex plane of the spectrum of the Hamiltonian is used \cite{BohGadMax97a}, which requires extending the quantum mathematical formalism of Hilbert spaces to that of rigged Hilbert spaces. We further conjecture that obtaining a Lindbladian evolution equation with just one nonselfadjoint Lindblad operator is also related to the existence of complex Hamiltonian eigenvalues and thus to micro-irreversibility \cite{BohGadMax97a}.\\
In author's opinion, a remarkable advantage of the approach presented above (apart from some mathematical benefits in solving some master equations \cite{SalSan02d,SalSan02b}) stems from the possibility of dealing with approximate (though necessary) descriptions such as the JCM, where the approximation itself carries an inherent decohering effect as a result of neglecting some degrees of freedom of the system. This effect has been expressed through the randomization of some internal parameters of the system Hamiltonian and contrasted with some experimental data \cite{SalSan02b}. The good agreement achieved invites us to look for new realistic situations in which we can further check the utility of this approach. A second advantage obtained from the involved mathematical formalism is the easy and fast generalization to nonmarkovian situations. However the approach is of phenomenological nature, consequently if one is interested in specific details depending e.g. on the concrete interaction between the system and the environment, it is compulsory to resort to the well-known ``tracing-out'' methods \cite{GiuJooKieKupStaZeh96a, BrePet02a}.\\
To sum up, in this work we have shown how the conjuction of stochastic methods and the spectral decomposition theorem can provide the analytical expression for the experimental behaviour of certain decohering quantum systems in a rather simple form. The main idea consists of identifying the pervasive effect of the surrounding environment upon a quantum system as a random perturbation of its evolution operator. Diverse decoherence functions can be obtained by adequately choosing the correlation properties of the random parameters. As a matter of fact it can be proved that when there exists correlation between these different parameters, then the system evolution is not completely positive, providing a further physical condition by which complete positivity is broken. It has been shown, as a noticeable example, how in a compound system the center of mass may suffer from decohering effects whereas the internal dynamics, i.e. the internal modes of the system are completely unaffected. The utility of this approach to deal with some approximate descriptions of quantum systems has been briefly discussed.

\appendix

\section{Computation of $\mathbb{E}\left[\exp\left(-\rmi\mathfrak{A}(t)\right)\right]$\label{Com}}

Here we include the computation of the stochastic average of the exponential of a \emph{random superoperator} $\mathfrak{A}(t)=\int_{0}^{t}\beta(s)\mathfrak{S}(s)d\mathcal{B}(s)$, where $\beta(s)$ is a deterministic function, $\mathfrak{S}(s)$ is an arbitrary superoperator and $\mathcal{B}(s)$ is a complex or real Brownian motion. By linearity of the expectation value, the computation reduces to find the $n$th order moment of $\mathfrak{A}(t)$. To do this, notice that it can be expressed through the Ito equation $d\mathfrak{A}(t)=\beta(t)\mathfrak{S}(t)d\mathcal{B}(t)$ and then applying Ito's formula to the function $\psi(\mathfrak{A}(t))=\mathfrak{A}^{n}(t)$. The $n$th order moment $\mathfrak{M}_{n}(t)\equiv\mathbb{E}[\mathfrak{A}^{n}(t)]$ then satisfies the recursive differential equation

\begin{equation}
d\mathfrak{M}_{n}(t)=\frac{n(n-1)}{2}\mathfrak{M}_{n-2}(t)\beta^{2}(t)\mathfrak{S}^{2}(t)d\langle\mathcal{B}\rangle_{t}\quad n\geq2
\end{equation}

Recalling that $\mathfrak{M}_{0}=1$ and $\mathfrak{M}_{1}=0$ one finally arrives at 

\begin{eqnarray}
\mathfrak{M}_{2n}(t)&=&\frac{(2n)!}{2^{n}n!}\mathfrak{K}^{n}(t)\\
\mathfrak{M}_{2n+1}(t)&=&0
\end{eqnarray}

\noindent where $\mathfrak{K}(t)\equiv\int_{0}^{t}\beta^{2}(s)\mathfrak{S}^{2}(s)d\langle B\rangle_{s}$. Substituting these relations in the series development of the exponential one immediately arrives at

\begin{equation}
\mathbb{E}[\exp\left(-\rmi\mathfrak{A}(t)\right)]=\exp\left(-\frac{\mathfrak{K}(t)}{2}\right)
\end{equation}

\section{Initially correlated environments might preclude completely positive system evolutions\label{InitCorr}}

Here we prove how the presence of quantum correlations inside the environment might prevent the system evolution from being completely positive, even when the system is not initially correlated with the environment. Let the whole system+environment be represented by a Hilbert space $\mathfrak{H}_{\rm S}\otimes\mathfrak{H}_{\rm 1}\otimes\mathfrak{H}_{\rm 2}$, where $\mathfrak{H}_{\rm S}$ corresponds to the system and $\mathfrak{H}_{\rm 1}\otimes\mathfrak{H}_{\rm 2}$ to the environment. Let the respective dimensions be $N_{\rm S}$, $N_{\rm 1}$ and $N_{\rm 2}$ and the generators of the group $\rm{SU}(N_{k})$ be denoted by $\sigma_{j}^{(k)}$, with $j=1,\dots,N_{k}$. The most general joint density operator can then be written as

\begin{eqnarray}
\rho_{\rm{S12}}&=&\frac{1}{N_{\rm S}N_{\rm 1}N_{\rm 2}}\left[\mathbb{I}_{\rm{S12}}+\alpha_{i}\sigma^{(\rm{S})}_{i}\otimes\mathbb{I}_{\rm 1}\otimes\mathbb{I}_{\rm 2}+\beta_{j}\mathbb{I}_{\rm S}\otimes\sigma^{(\rm 1)}_{j}\otimes\mathbb{I}_{\rm 2}+\gamma_{k}\mathbb{I}_{\rm S}\otimes\mathbb{I}_{\rm 1}\otimes\sigma^{(\rm 2)}_{k}+\right.\nonumber\\
&\lo+&\delta_{ij}\sigma^{(\rm S)}_{i}\otimes\sigma^{(\rm 1)}\otimes\mathbb{I}_{\rm 2}+\epsilon_{ik}\sigma^{(\rm S)}_{i}\otimes\mathbb{I}_{\rm 1}\otimes\sigma^{(\rm 2)}_{k}+\eta_{jk}\mathbb{I}_{\rm S}\otimes\sigma^{(\rm 1)}_{j}\otimes\sigma^{(\rm 2)}_{k}+\nonumber\\
&\lo+&\left.\nu_{ijk}\sigma^{(\rm S)}_{i}\otimes\sigma^{(\rm 1)}_{j}\otimes\sigma^{(\rm 2)}_{k}\right]
\end{eqnarray}

Then denoting $\rho_{k}(t)=\rm{Tr}_{lm}\left(U_{\rm{S12}}(t)\rho_{\rm{S12}}(0)U_{\rm{S12}}^{\dagger}(t)\right)$ with $l\neq k\neq m$, it is elementary to prove that $\rho_{\rm{S12}}$ can be rewritten as 

\begin{eqnarray} 
\rho_{\rm{S12}}&=&\rho_{\rm S}\otimes\rho_{\rm 1}\otimes\rho_{\rm 2}+\delta'_{ij}\sigma^{(\rm S)}_{i}\otimes\sigma^{(\rm 1)}\otimes\mathbb{I}_{\rm 2}+\epsilon'_{ik}\sigma^{(\rm S)}_{i}\otimes\mathbb{I}_{\rm 1}\otimes\sigma^{(\rm 2)}_{k}+\nonumber\\
&+&\eta'_{jk}\mathbb{I}_{\rm S}\otimes\sigma^{(\rm 1)}_{j}\otimes\sigma^{(\rm 2)}_{k}+\nu'_{ijk}\sigma^{(\rm S)}_{i}\otimes\sigma^{(\rm 1)}_{j}\otimes\sigma^{(\rm 2)}_{k}
\end{eqnarray}

\noindent where 

\numparts
\begin{eqnarray}
\delta'_{ij}\equiv\frac{1}{N_{\rm S}N_{\rm 1}N_{\rm 2}}\left(\delta_{ij}-\alpha_{i}\beta_{j}\right)&\quad&\epsilon'_{ij}\equiv\frac{1}{N_{\rm S}N_{\rm 1}N_{\rm 2}}\left(\epsilon_{ij}-\alpha_{i}\gamma_{j}\right)\\
\eta'_{ij}\equiv\frac{1}{N_{\rm S}N_{\rm 1}N_{\rm 2}}\left(\eta_{ij}-\beta_{i}\gamma_{j}\right)&\quad&\nu'_{ijk}\equiv\frac{1}{N_{\rm S}N_{\rm 1}N_{\rm 2}}\left(\nu_{ijk}-\alpha_{i}\beta_{j}\gamma_{k}\right)
\end{eqnarray}
\endnumparts

From this and after well-known arrangements \cite{SteBuz01a}, it is easy to arrive at

\begin{eqnarray}
\fl\rho_{\rm{S}}(t)&=&\sum_{a b}\sum_{a' b'}M_{ab,a'b'}(t)\rho_{\rm{S}}(0)M_{ab,a'b'}^{\dagger}+\nonumber\\
\lo+&&\delta'_{ij}\rm{Tr}_{\rm{12}}\left(U(t)\sigma_{i}^{(\rm{S})}\otimes\sigma^{(\rm{1})}_{j}\otimes\mathbb{I}_{\rm{2}}U^{\dagger}(t)\right)+\epsilon'_{ik}\rm{Tr}_{\rm{12}}\left(U(t)\sigma^{(\rm{S})}_{i}\otimes\mathbb{I}_{\rm{1}}\otimes\sigma^{(\rm{2})}_{k}U^{\dagger}(t)\right)+\nonumber\\
\lo+&&\eta'_{ij}\rm{Tr}_{\rm{12}}\left(U(t)\mathbb{I}_{\rm{S}}\otimes\sigma^{(\rm{1})}_{i}\otimes\sigma^{(\rm{2})}_{k}U^{\dagger}(t)\right)+\nu'_{ijk}\rm{Tr}_{\rm{12}}\left(U(t)\sigma^{(\rm{S})}_{i}\otimes\sigma^{(\rm{1})}_{j}\otimes\sigma^{(\rm{2})}_ {k}U^{\dagger}(t)\right)\nonumber\\\label{RedDensMat}
\end{eqnarray}

From \eref{RedDensMat} it is clear that even when the system and the environment are not initially correlated ($\delta'_{ij}=\epsilon'_{ik}=\nu'_{ijk}=0$), the initial correlations inside the environment $\eta'_{jk}\neq 0$ might introduce a non-Kraus term in the evolved reduced density matrix of the system, thus precluding complete positivity. An example has been provided in the text using different methods.\\
However note that when the coefficients satisfy

\numparts
\begin{eqnarray}
\alpha_{i}=\bar{\alpha}_{i}&\beta_{j}=\bar{\beta}_{j}&\gamma_{k}=\bar{\gamma}_{k}\\
\delta_{ij}=\bar{\alpha}_{i}\bar{\beta}_{j}&\quad\epsilon_{ik}=\bar{\alpha}_{i}\bar{\gamma}_{k}&\quad\eta_{jk}=\bar{\eta}_{jk}\\
&\nu_{ijk}=\bar{\alpha}_{i}\bar{\eta}_{jk}
\end{eqnarray}
\endnumparts

\noindent for arbitrary ``barred'' quantities, the joint density operator can then be factorized as $\rho_{\rm{S12}}=\rho_{\rm{S}}\otimes\rho_{\rm{12}}$ and the reduced dynamics is then completely positive as expected. Another sufficient condition to get a completely positive evolution even in the case of initial correlation arises when the dynamics is not entangling \cite{Bus02a}, i.e. when the evolution operator can be written as $U_{\rm{S12}}(t)=U_{\rm{S}}(t)\otimes U_{\rm{1}}(t)\otimes U_{\rm{2}}(t)$ (see \cite{SalSan02f} for details).
\ack
One of us (D.S.)  acknowledges the support of Madrid Education Council under grant BOCAM 20-08-1999. 

\section*{References}


\end{document}